\documentclass[superscriptaddress,prl,floatfix,twocolumn,amsmath,amssymb,aps,longbibliography, footinbib]{revtex4-2}
\usepackage{graphicx}
\usepackage{amsmath}
\usepackage{amssymb}
\usepackage{color}
\usepackage{multirow}
\usepackage{bm}
\usepackage{subfigure}
\usepackage{tabularx}
\usepackage{hhline}  
\newcolumntype{d}[1]{D{.}{.}{#1}}

\DeclareMathOperator{\Tr}{Tr}
\newcommand{\dpar}[2]{\frac{\partial #1}{\partial #2}}
\newcommand{\bracket}[1]{\langle #1 \rangle}
\newcommand{\ket}[1]{|#1\rangle}

\begin{document}
	\title{Gate-tunable phonon magnetic moment in bilayer graphene}
	\author{Xiao-Wei Zhang}
	\affiliation{Department of Materials Science and Engineering, University of Washington, Seattle, WA 98195, USA}
	\author{Yafei Ren}
	\affiliation{Department of Materials Science and Engineering, University of Washington, Seattle, WA 98195, USA}
	\author{Chong Wang}
	\affiliation{Department of Materials Science and Engineering, University of Washington, Seattle, WA 98195, USA}
	\author{Ting Cao}
	\email{tingcao@uw.edu}
	\affiliation{Department of Materials Science and Engineering, University of Washington, Seattle, WA 98195, USA}
	\author{Di Xiao}
	\email{dixiao@uw.edu}
	\affiliation{Department of Materials Science and Engineering, University of Washington, Seattle, WA 98195, USA}
	\affiliation{Department of Physics, University of Washington, Seattle, WA 98195, USA}
	\date{\today}
\begin{abstract}
We develop a first-principles quantum scheme to calculate the phonon magnetic moment in solids.  As a showcase example, we apply our method to study gated bilayer graphene, a material with strong covalent bonds.  According to the classical theory based on the Born effective charge, the phonon magnetic moment in this system should vanish, yet our quantum mechanical calculations find significant phonon magnetic moments. Furthermore, the magnetic moment is highly tunable by changing the gate voltage.  Our results firmly establish the necessity of the quantum mechanical treatment, and identify small-gap covalent materials as a promising platform for studying tunable phonon magnetic moment.
\end{abstract}

\maketitle

Chiral phonons, characterized by circular motions of ions~\footnote{Here we use ``chiral phonons'' to refer to circularly rotating phonon modes. Strictly speaking, circularly rotating phonon modes that are not propagating do not have true chirality.}, have received intense recent interest due to their roles in various quantum phenomena, including the phonon Hall effect~\cite{grissonnanche2019giant,li2020phonon}, the Einstein-de Haas effect~\cite{zhang2014angular,dornes2019ultrafast}, valleytronics~\cite{zhang2015chiral,zhu2018observation}, exciton-phonon replica~\cite{li2019emerging,he2020valley}, and magnon excitations by pumping optical phonons~\cite{nova2017effective}.  An important property of chiral phonons is that they can carry an orbital magnetic moment, allowing for the prospect of dynamic generation of magnetization even in non-magnetic materials~\cite{juraschek2017dynamical}. In fact, several experiments have reported the observation of phonon magnetic moments~\cite{cheng2020large,baydin2022magnetic,hernandez2022chiral,basini2022terahertz}. 
On the theory side, Juraschek~\textit{et al.}~have calculated the phonon magnetic moment using the Born effective charge of ions in circular motion~\cite{juraschek2019orbital}.  This classical theory naturally points to ionic materials as the most likely material class with large phonon magnetic moments.

However, the classical theory does not explicitly account for the quantum nature of electrons and thus will fail in covalent materials.  To this end, Ceresoli and Tosatti have proposed a quantum theory of rotational $g$-factor in terms of the molecular Berry phase~\cite{ceresoli2002berry,ceresoli2002PRL,zabalo2022rotational}.  More recently, an alternative quantum mechanical treatment based on adiabatic pumping has identified two distinct contributions to the phonon magnetic moment ~\cite{dong2018geometrodynamics,trifunovic2019geometric,xiao2021adiabatically,ren2021phonon}: the topological contribution, which is the quantum extension of the classical theory and is related to the momentum-resolved Born effective charge; and the perturbative contribution, which results from the adiabatic correction to the electronic wave functions induced by phonons.  These theoretical advances raise the question of whether phonon magnetic moment can be significant in covalent materials, which would provide a new playground for dynamic magnetization generation. 

In this Letter, we develop a first-principles quantum scheme based on finite-difference and gauge-covariant techniques to calculate phonon magnetic moment in a wide range of materials from very ionic to very covalent.  In these materials, we find significant difference between the classical and the quantum theories.  Within the quantum theory, the perturbative contribution is as large as, and sometimes dominates over, the topological contribution.  As a showcase example, we apply our method to study gated bilayer graphene, a material with strong covalent bonds.  According to the classical theory, the phonon magnetic moment in this system should vanish, yet our quantum mechanical calculations find significant phonon magnetic moments. Furthermore, the phonon magnetic moment is highly tunable by changing the gate voltage. Under an external out-of-plane electric field of 25~mV/\AA, the magnetic moment of a chiral shear mode can reach $\sim$ 0.01 $\mu_{\text{B}}$, representing the largest phonon magnetic moment among existing theoretical predictions.  Our results firmly establish the necessity of the quantum mechanical treatment, and identify small-gap covalent materials as a promising platform for studying tunable phonon magnetic moment.

\begin{figure}[t]
\includegraphics[width=1.0\columnwidth]{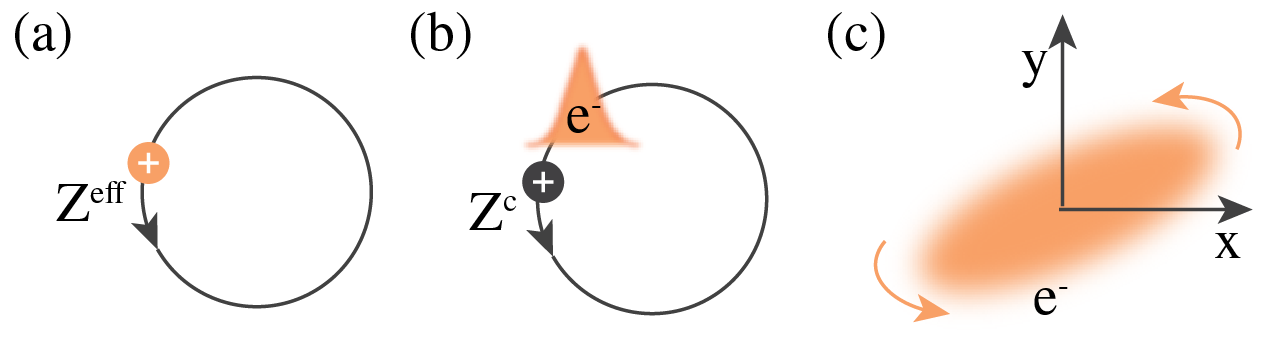}
	\caption{The physical pictures of phonon magnetic moments from (a) the classical theory, and the quantum theory which includes (b) the topological part and (c) the perturbative part.  $Z^{\text{eff}}$ and $Z^{\text{c}}$ denote the Born effective charge (electron plus nuclear charge) and the nuclear charge, respectively.  In the classical picture, the motion of the electron and nucleus is described by a circulating Born effective charge.  In the quantum picture, the motion of nuclei is classical while the motion of electrons is described by a wave packet.  In (b), the wave packet has center-of-mass orbital motion driven by circular motions of the nucleus.  In (c), the wave packet has a self-rotation due to a time-dependent anisotropic potential.}
	\label{fig1}
\end{figure}

We begin with a brief review of the classical and quantum theories of phonon magnetic moment. The phonon magnetic moment includes the electron's and nuclear contributions.
In the classical theory, the material is treated as a collection of ions each having mass $M_i$, with their charge given by the Born effective charge $Z^\text{eff}_i$ (electron plus nuclear charge).  The circular motion of ions is illustrated in Fig.~\ref{fig1}(a).  Then the total magnetic moment (electron plus nucleus) of a phonon mode $\nu$ is written as~\cite{juraschek2019orbital}
\begin{equation} \label{born}
    \bm M^{\text{clas}}(\nu) = i\hbar \sum_i \gamma_i \bm\xi_i^{(\nu)} \times \bm\xi_i^{(\nu)\ast} \;,
\end{equation}
where $\bm\xi_i^{(\nu)}$ is the normalized phonon eigenvector component of the $i$th ion in a phonon mode $\nu$ ($\sum_i |\bm\xi_i^{(\nu)}|^2 = 1$), and $\gamma_i \equiv eZ^{\text{eff}}_i/(2M_i)$ is the gyromagnetic ratio.  In ionic materials, the cations and the anions have different masses. Although they contribute opposite moments,  the net magnetic moment is nonzero.  On the other hand, small phonon magnetic moment in covalent materials is expected according to Eq.~\eqref{born}.

In the quantum theory, the nuclear contribution has the same form as the classical theory, while the electron's contribution to the phonon magnetic moment is derived under the adiabatic condition, i.e., the phonon frequency must be smaller than the band gap.  As we mentioned earlier, there are two contributions to the phonon magnetic moment.  The topological contribution, denoted by $\bm M^\text{top}$, can be expressed as a Chern-Simons integral in the parameter space spanned by $\bm k$ and {time~\cite{trifunovic2019geometric}. 
In the limit of small phonon displacement, one can expand $\bm M^\text{top}$ and write it as an integral of a second Chern form~\cite{ren2021phonon}.  This expression has the advantage that the integrand contains only gauge-invariant quantities, and is therefore more suitable for numerical implementation.  For a phonon mode $\nu$ at the $\Gamma$ point, $\bm M^\text{top}$ is given by
\begin{equation} \label{top}
\begin{split}
M_z^\text{top}(\nu) &= \frac{e\hbar}{4 N_k}
\sum_{\bm k} \sum_{i\delta, j\gamma} \Bigl[ 
- i\xi_{i\delta}^{(\nu)} \xi_{j\gamma}^{(\nu)\ast} 
+ i\xi_{i\delta}^{(\nu)\ast} \xi_{j\gamma}^{(\nu)}
\Bigr] \\  
& \quad \times \frac{1}{\sqrt{M_iM_j}} \Tr\Omega_{xy\gamma\delta}^{kku_{j}u_{i}}
\Big|_{\bm R^0} \;,
\end{split}
\end{equation}
where 
\begin{equation} \label{chern}
\Omega_{xy\gamma\delta}^{kku_{j}u_{i}} \equiv
\Omega_{xy}^{kk} \Omega_{\gamma\delta}^{u_ju_i}
+\Omega_{y\gamma}^{ku_j}\Omega_{x\delta}^{ku_i}
-\Omega_{x\gamma}^{ku_j}\Omega_{y\delta}^{ku_i}
\end{equation}
is the second Chern form.  In the above equations, $N_k$ is the number of $\bm k$-grid points in the Brillouin zone, $i$ and $j$ label the atoms in the unit cell, $\gamma$ and $\delta$ are Cartesian components, and the whole expression is evaluated with respect to the equilibrium position $\bm R^0$ of ions. The central quantities are the non-Abelian Berry curvatures, $\Omega_{xy}^{kk}$, $\Omega_{\gamma\delta}^{u_ju_i}$ and $\Omega_{x\gamma}^{ku_j}$, defined in the parameter space of momentum $\bm k$ and displacement $\bm u_i$.  For example, $\Omega_{x\delta}^{ku_j} \equiv \partial_{k_x}A_{u_{j\delta}} - \partial_{u_{j\delta}}A_{k_x} - i[A_{k_x}, A_{u_{j\delta}}]$, where $A_{k_x}$ is the Berry connection with its matrix element given by $A_{k_x, mn}\equiv\bracket{\phi_{m\bm k}|i\partial_{k_x}|\phi_{n\bm k}}$ and $\ket{\phi_{n\bm k}}$ is the periodic part of the Bloch function of the $n$th band. The trace is taken over occupied bands. The definition of other Berry connections can be found in the Supplementary Materials~\cite{supp}.
According to Ref.~\cite{ren2021phonon}, the topological contribution originates from the center-of-mass motion of the electron wave packet and therefore can be regarded as a quantum extension of the classical theory [Fig. ~\ref{fig1}(b)].  Indeed, as we show later, in the extreme ionic case $\bm M^\text{top}$ coincides with the value obtained from the classical theory.

In addition to the topological contribution, phonons also cause an adiabatic correction to the electronic wave function, giving rise to a perturbative contribution, $\bm M^\text{pert}$, to the phonon magnetic moment.  According to Ref. ~\cite{trifunovic2019geometric}
\begin{equation} \label{ntop}
\begin{split}
   M^\text{pert}_z(\nu) &= \frac{e\hbar}{4} 
   \Re \sum_{i\delta,j\gamma} \Bigl[i\xi_{i\delta}^{(\nu)} \xi_{j\gamma}^{(\nu)\ast} - i \xi_{i\delta}^{(\nu)\ast} \xi_{j\gamma}^{(\nu)}\Bigr] \\ 
   &\quad\times \frac{1}{\sqrt{M_iM_j}}
   \Bigl[\dpar{F(u_{i\delta})}{u_{j\gamma}} 
   - \dpar{F(u_{j\gamma})}{u_{i\delta} }\Bigr]_{\bm R^0} \;,
\end{split}
\end{equation}
where
\begin{equation} \label{F}
\begin{split}
F(u_{i\delta}) &= \frac{1}{N_{k}}
\sum_{\bm k} \sum_n^\text{occ} \sum_{n'm}^\text{unocc} 
\frac{\bracket{\dpar{\phi_{n\bm k}}{u_{i\delta}}|\phi_{n'\bm k}}
v_{n'm\bm k} \times v_{mn\bm k}}
{(E_{n\bm k}-E_{n'\bm k})(E_{n\bm k}-E_{m\bm k})} \\ 
&\quad -\frac{1}{N_{k}}
\sum_{\bm k}\sum_{mn}^{\text{occ}}\sum_{n'}^{\text{unocc}}
\frac{\bracket{\dpar{\phi_{n\bm k}}{u_{i\delta}}|\phi_{n'\bm k}}
v_{n'm\bm k} \times v_{mn\bm k}}
{(E_{n\bm k}-E_{n'\bm k})(E_{m\bm k}-E_{n'\bm k})} \;.
\end{split}
\end{equation}
Here, $v_{mn\bm k}$ is the velocity matrix element, and ``occ'' and ``unocc'' denote occupied and unoccupied bands, respectively.  $\bm M^\text{pert}$ originates from the self-rotation of the electron wave packet forced by the rotation of the anisotropic potential due to chiral phonons, as schematically shown in Fig.~\ref{fig1}(c)~\cite{trifunovic2019geometric}. $\bm M^\text{pert}$ has no classical counterpart.

\begin{figure*}[t]
	\includegraphics[width=0.8\linewidth]{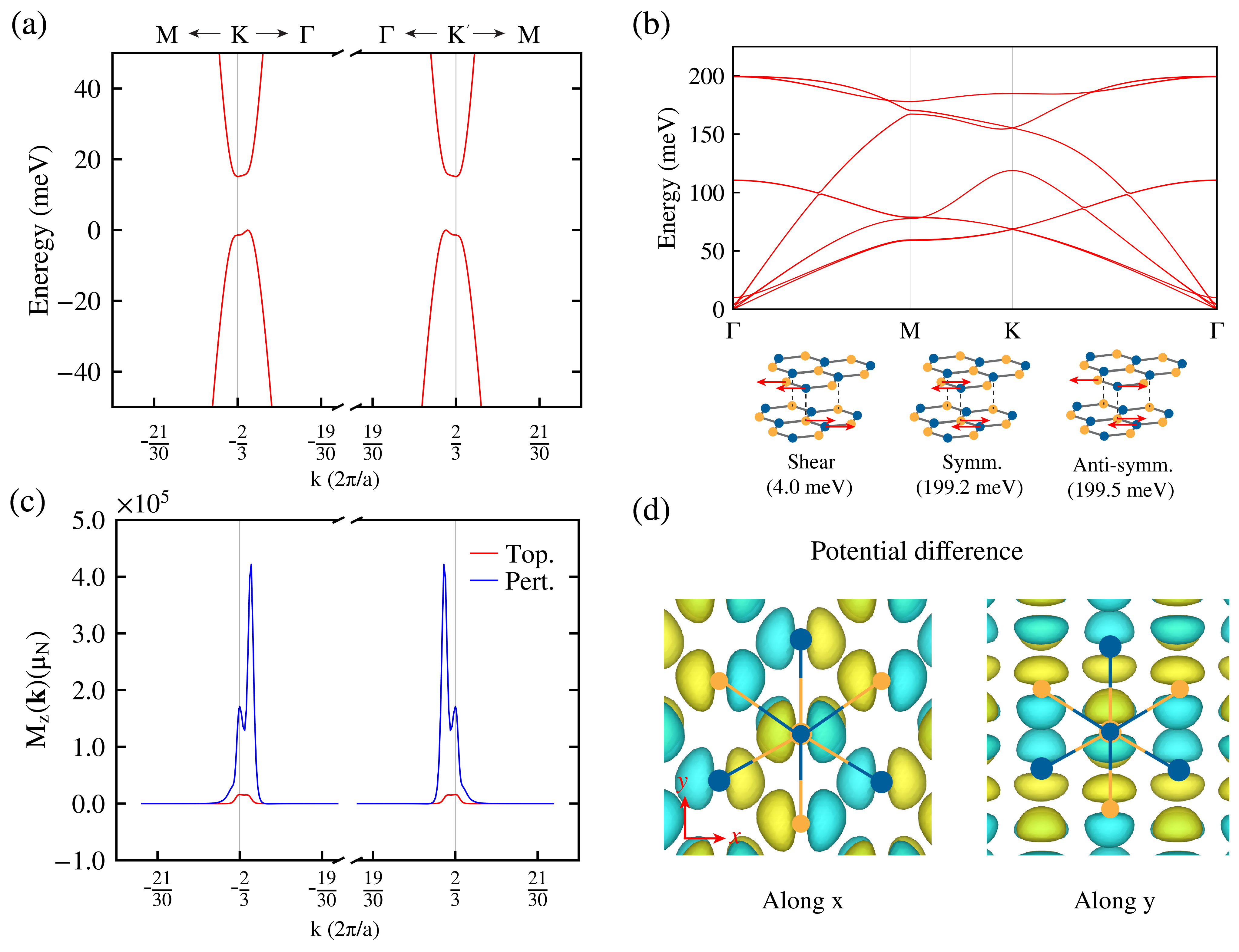}
	\caption{ (a) Electron's Kohn-Sham band structure of bilayer graphene under an out-of-plane electric field $E = 25$ mV/\AA.  (b) Upper: the phonon band structure of bilayer graphene without the electric field.  Bottom: the vibrational illustrations of doubly-degenerate phonon modes for the shear, symmetric and anti-symmetric modes at the $\Gamma$ point (in ascending energy). The numbers in the bracket denote the phonon frequencies. (c) Momentum resolved electron's topological and perturbative contributions to the magnetic moment of the chiral shear mode at $E = 25$ mV/\AA. $M_z$ ($\bm{k}$) is defined at each $\bm{k}$-point without being divided by the number of $\bm{k}$-points $N_{k}$. The topological and perturbative contributions to the magnetic moments in Table~\ref{tab} are calculated by averaging $M_z$ ($\bm{k}$) in the BZ.
 The unit of magnetic moments is $\mu_{\text{N}}=e\hbar/2M_{\text{N}}$, where $M_{\text{N}}$ is the mass of proton. 
 (d) The first-principles Kohn-Sham self-consistent potential difference between the distorted and equilibrium structure at $E= 25$ mV/\AA.  The left and right panels correspond to the structures where the two layers are oppositely displaced by 0.01 $\text{\AA}$ along the $x$- and $y$-directions, respectively.  The yellow and blue colors denote the positive and negative values, respectively.  The isosurface level is $\pm$ 0.05 V. Atomic positions are labeled by colored dots. 
	}
	\label{fig2}
\end{figure*}

To implement the quantum theory in first-principles calculations, we adopt the technique of gauge-covariant derivatives, which has proven to be robust against band crossing ~\cite{sai2002theory,souza2004dynamics,wang2006ab}.  As a result,  we can use an efficient finite-difference formula to calculate various derivatives and matrix elements in Eq. \eqref{top} - Eq. \eqref{F}.  In our calculations, the Quantum Espresso package~\cite{giannozzi2009quantum} is employed, and the ionic potential is treated by the optimized norm-conserving Vanderbilt pseudopotential (ONCVPSP)~\cite{hamann2013optimized}. Our method can be implemented in other DFT software packages as well, as it only needs wave functions, band energies, and velocity matrix elements from the DFT.  Details of our numerical implementation can be found in the Supplementary Material~\cite{supp}.  As a first benchmark, we have calculated the rotational $g$ factor of H$_2$ molecule.  Our result ($g = 0.8989)$ agrees well with the experimental value ($g = 0.8787$)~\cite{ramsey1940rotational}.

Having established the validity of our method, next we focus on Bernal-stacked bilayer graphene, a material with strong covalent bonds.  Since the spin-orbit coupling is negligible in graphene, there is no spin contribution to the phonon magnetic moments~\cite{hamada2020conversion}.  An out-of-plane electric field is applied to open a band gap larger than the phonon energy such that the adiabatic condition is satisfied.  We find that a dense $k$-point sampling is needed to converge the self-consistent calculations and resolve the band dispersions near the K and K$^{\prime}$ valleys [see Fig.~\ref{fig2}(a)].  We have chosen a $k$-space grid from $30\times 30 \times 1$ to $120\times 120 \times 1$, depending on the electric field strength. Computational details and convergence tests are included in Section II of the Supplementary Materials~\cite{supp}.

\begin{table}[t]
\centering
\bgroup
\def\arraystretch{1.3}
\caption{\label{tab}
  Comparison between phonon magnetic moments calculated from the classical and quantum theories, $M^{\text{clas}}=M^{\text{Born}}+M^{\text{nucl}}$ and $M^{\text{quan}}=M^{\text{top}}+M^{\text{pert}}+M^{\text{nucl}}$, where ``Born'' denotes the electrons' contribution in the classical theory and ``Nucl.'' denotes the nuclear contribution. The unit of magnetic moments is $\mu_{\text{N}}=e\hbar/2M_{\text{N}}$, where $M_{\text{N}}$ is the mass of proton. The numbers in the bracket denote phonon frequencies in meV.
     }
  \begin{tabularx}{1\columnwidth}{@{\extracolsep{\fill}} lcccccc}
    \hline\hline
      \multirow{2}*{} & \multicolumn{3}{c}{Electron}
    & \multicolumn{1}{c}{Nucl.} & \multicolumn{1}{c}{Quan.} & \multicolumn{1}{c}{Clas.}    \\ 
    & Top. & Pert. & Born &  & &    \\ 
    \hline 
     Shear$^\text{a}$   & 0.631 & 20.714 & -0.333 & 0.333 & 21.678 & $\sim$10$^{-4}$ \\ 
     (0.025 V/$\text{\AA}$, 4.0) \\ 
     Shear$^\text{a}$   & -0.286 & 0.119 & -0.333 & 0.333 & 0.177 & $\sim$10$^{-4}$ \\ 
     (0.3 V/$\text{\AA}$, 4.0) \\ 
      Symm.$^\text{a}$  & 0.601 & -19.732 &  0.338 & -0.333 &  -19.464 & 0.005 \\
      (0.5 V/$\text{\AA}$, 199.2) \\
     Anti-symm.$^\text{a}$ & 1.974 & 5.324 & 0.338 & -0.333 & 6.965 & 0.005 \\
     (0.5 V/$\text{\AA}$, 199.5) \\ 
     Si (63.5) & 0.004  & -0.022 & 0.143 & -0.142 & -0.160 & $\sim$ 10$^{-4}$ \\ 
     Blue P$^\text{b}$ (52.4) & 0.016 & -0.024 & 0.169 & -0.169 & -0.177 & $\sim $ 10$^{-6}$ \\ 
     MoS$_{2}^\text{b}$ (46.4) & 0.171 & -0.137 & 0.168 & -0.171 & -0.137 & -0.002  \\ 
     CsH (43.1) & -2.096  & 0.860 &  -2.096 & 0.985 & -0.251 & -1.111  \\ 
     \hline
\hline
\multicolumn{7}{l}{$^\text{a}$bilayer graphene; $^\text{b}$monolayer}
\end{tabularx}
\egroup
\end{table}

Figure~\ref{fig2}(b) shows the phonon spectrum and phonon modes of bilayer graphene without an external electric field.  At the $\Gamma$ point, there are three sets of doubly-degenerate in-plane phonon modes, including the shear (4.0 meV), the symmetric (199.2 meV), and the antisymmetric (199.5 meV) modes.  The shear and symmetric modes are Raman-active whereas the antisymmetric mode is infrared-active.  Within each degenerate pair, a left and a right circularly-polarized phonon modes can be constructed~\footnote{Ref.~\cite{ren2021phonon} considered the magnetic moment of the $K$-point phonons in gapped monolayer graphene. Here we consider the $\Gamma$-point phonons in a gated bilayer graphene}}. We will consider the magnetic moment of a specific circular polarization. Modes with opposite circular polarization have opposite magnetic moments.
We find that the electric field has negligible effects on the phonon frequencies and the eigenvectors of the shear mode. However, at large electric fields, e.g. $E = 0.5$ V/\AA, there is a slight mixing between the eigenvectors of the symmetric and antisymmetric modes.

We first consider the shear mode at 4.0 meV.  To satisfy the adiabatic condition, a small electric field of 25 mV/\AA\ has been applied to open a band gap of 15 meV as shown in Fig.~\ref{fig2}(a).  Figure~\ref{fig2}(c) shows the distribution of the topological and the perturbative contributions to the magnetic moment along the high-symmetry line in the Brillouin zone.  Both contributions are sharply centered around the K and K$^{\prime}$ point and the double-peak feature is caused by the Mexican-hat band dispersion.  After summing over both contributions over the Brillouin zone, we obtain the result from the quantum theory as shown in Table~\ref{tab}.  The total magnetic moment of the shear mode is 21.678 $\mu_\text{N}$ ($\sim$ 0.012 $\mu_\text{B}$), where $\mu_\text{N}$ and $\mu_{\text{B}}$ are nuclear magneton and Bohr magneton, respectively.  This is in fact the largest value among existing theoretical predictions of phonon magnetic moment. We also note that the perturbative contribution dominates over the topological contribution.  In comparison, we have also calculated the phonon magnetic moment using the classical theory.  The Born effective charge becomes finite upon applying an electric field; however, the classical phonon magnetic moment still vanishes because the four carbon atoms have the same radii of the circular motion and the summation of Born effective charges of the four carbon atoms is zero.

To gain a more intuitive understanding of the magnetic moment, in Fig.~\ref{fig2}(d) we plot the difference of the Kohn-Sham self-consistent potential between the distorted (move the first layers by 0.01 $\text{\AA}$ along the $x$ or $y$ direction) and the equilibrium structure.  It can be seen that the differential potential becomes highly anisotropic in response to the displacement.  When the two layers shift against each other in a circular motion as in the chiral shear mode, the resulting potential variation has two effects.  First, because the potential minimum changes, the electron wave packet will adjust its center-of-mass position to follow the ions, which corresponds to the topological contribution~\cite{ren2021phonon, trifunovic2019geometric}.  Second, the potential distribution is anisotropic as can be seen by comparing the potential shapes along the $x$ and $y$ directions. This will cause a self-rotation of the electron wave packet and give rise to a magnetic moment~\cite{trifunovic2019geometric}.  This scenario corresponds to the perturbative contribution. 

\begin{figure}[t]
	\includegraphics[width=1.0\columnwidth]{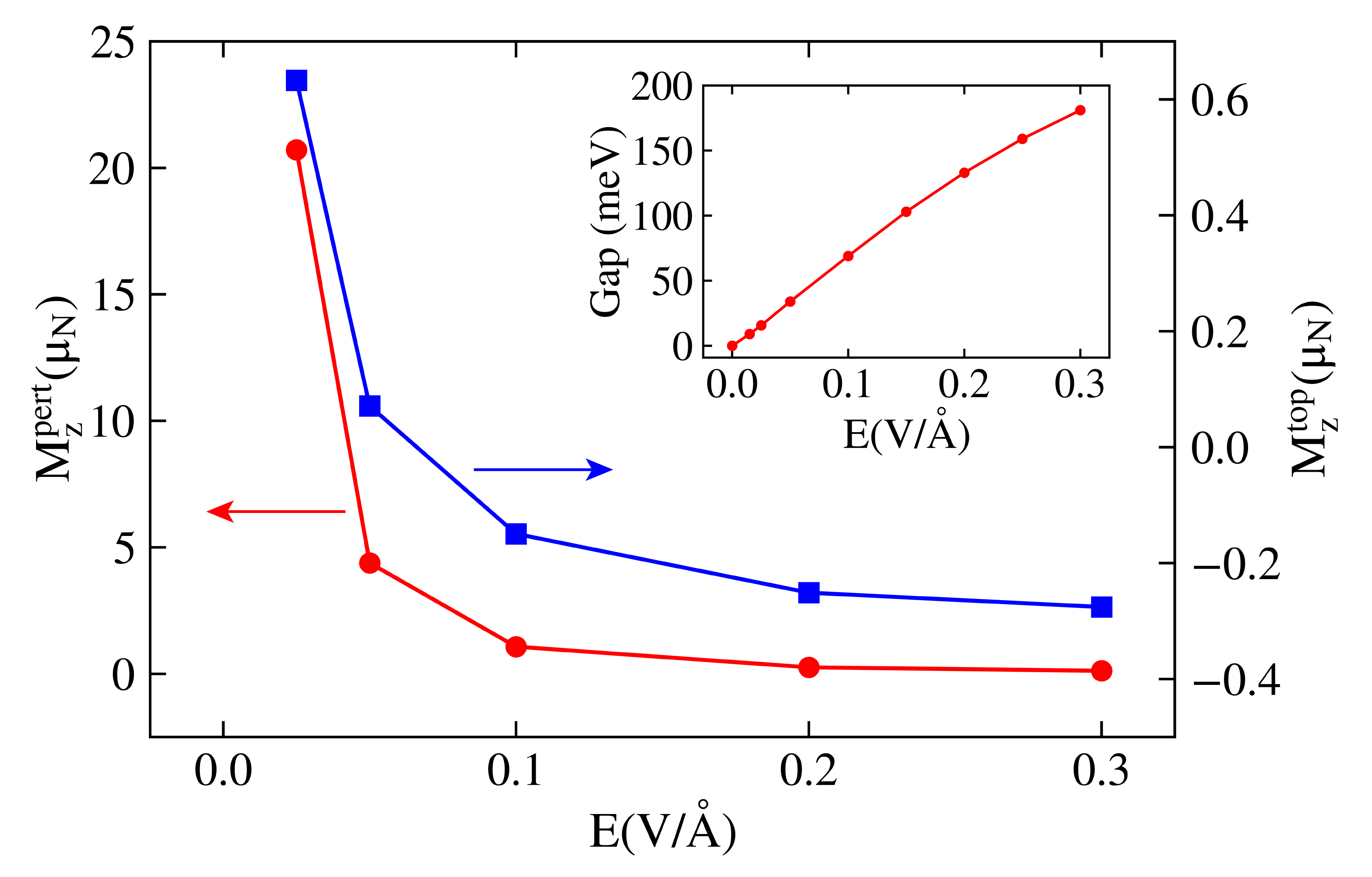}
	\caption{ The electric-field dependence of the perturbative (red circles, left axis) and the topological (blue squares, right axis) contributions  of the magnetic moment for the shear mode. Lines are guides to the eye. Inset: the electric-field dependence of band gap from DFT-LDA calculations. 
	}
	\label{fig3}
\end{figure}

The phonon magnetic moment can be tuned by changing the electric field, although the phonon modes remain virtually unchanged.  Figure~\ref{fig3} shows the field dependence of the topological and perturbative contributions to the magnetic moment of the shear mode.  In our calculations, the electric field is in the experimentally accessible range of 25 to 300 mV/\AA, leading to band gaps from 15 meV to 180 meV (see the inset of Fig.~\ref{fig3}). As the electric field increases, the topological contribution changes sign due to gate-induced changes of Berry curvature distribution in the BZ~\cite{supp}.
On the other hand, the perturbative contribution is, approximately, inversely proportional to the band gap squared [see Eq.~\eqref{F}], and therefore continuously drops to zero at large electric fields. 
Another feature of the phonon magnetic moments is that they are independent of the direction of the out-of-plane electric field. This is because, under inversion operation, the electric field flips sign, whereas the phonon modes and angular momenta are invariant. 

In addition to the shear mode, we have also calculated the phonon magnetic moment of the symmetric (199.2 meV) and the antisymmetric (199.5 meV) modes.  To satisfy the adiabatic condition, we have applied an electric field of $E = 0.5$ V/\AA, which opens a band gap of 240 meV.  We find the magnetic moment of the symmetric mode is $-19.464$ $\mu_\text{N}$ ($\sim-0.011$ $\mu_\text{B}$), and the magnetic moment of the anti-symmetric mode is 6.965 $\mu_\text{N}$ ($\sim$0.004 $\mu_\text{B}$) [Table~\ref{tab}].  Note that since the DFT-LDA usually underestimates the band gap, in experiments an electric field of 0.3 V/\AA\ is sufficient to satisfy the adiabatic condition~\cite{zhang2009direct}. 

Our newly developed numerical method allows us to calculate the phonon magnetic moments for a wide range of materials listed in Table~\ref{tab}.  We begin with the extreme covalent materials, namely, bulk silicon and monolayer blue phosphorus. The Born effective charges in both materials are zero, but the quantum theory predicts non-zero phonon magnetic moments~\footnote{We note that the magnetic moment is invariant under spatial inversion, therefore the phonon magnetic moment can be nonzero even in covalent materials with inversion symmetry such as silicon and blue phosphorus listed in Table~\ref{tab}}. Moving on to monolayer MoS$_2$, we see that the topological contribution is getting closer to the result based on the electron's Born effective charge, and nearly cancels the nuclear contribution.  As a result, the total phonon magnetic moment is mostly given by perturbative contribution. This feature is entirely missing from the classical theory. Finally, in the extreme ionic case, e.g., CsH, the topological contribution becomes significant and identical to that from Born effective charge.  However, the perturbative contribution still accounts for a sizable part of the total phonon magnetic moment.  These results clearly demonstrate the necessity of using the quantum theory to calculate the phonon magnetic moment.

In summary, we have developed a first-principles scheme to implement the quantum theory of phonon magnetic moment for realistic materials.  We find that the phonon magnetic moment can be significant in small-gap covalent materials, where the classical theory based on the Born effective charge fails completely.  In particular, bilayer graphene serves as an ideal platform to test the current adiabatic theory. The shear modes in our calculation are Raman active, and their magnetic moments can be detected by Raman scattering under magnetic fields~\cite{schaack1976observation, schaack1977magnetic}. Our findings can also be extended to trilayer graphene, where the shear modes are infrared active. Their magnetic moments can be measured by the adsorption of THz light under magnetic fields~\cite{cheng2020large}.  Intense THz light can also be used to generate coherent chiral phonons for the manipulation of the magnetic properties of trilayer graphene dynamically~\cite{shin2018phonon,basini2022terahertz}. The boundary current from the dynamically induced magnetization is detectable by the NV center-based quantum sensor~\cite{schirhagl2014nitrogen}.  Our results predict a new class of materials to realize tunable phonon magnetic moment and call for more experimental investigations.

We thank Qian Niu for stimulating discussions.  This work is supported by DOE Award No.~DE-SC0012509.  This work was facilitated through the use of advanced computational, storage, and networking infrastructure provided by the Hyak supercomputer system and funded by the University of Washington Molecular Engineering Materials Center at the University of Washington (DMR-1719797).


\begin{thebibliography}{38}%
\makeatletter
\providecommand \@ifxundefined [1]{%
 \@ifx{#1\undefined}
}%
\providecommand \@ifnum [1]{%
 \ifnum #1\expandafter \@firstoftwo
 \else \expandafter \@secondoftwo
 \fi
}%
\providecommand \@ifx [1]{%
 \ifx #1\expandafter \@firstoftwo
 \else \expandafter \@secondoftwo
 \fi
}%
\providecommand \natexlab [1]{#1}%
\providecommand \enquote  [1]{``#1''}%
\providecommand \bibnamefont  [1]{#1}%
\providecommand \bibfnamefont [1]{#1}%
\providecommand \citenamefont [1]{#1}%
\providecommand \href@noop [0]{\@secondoftwo}%
\providecommand \href [0]{\begingroup \@sanitize@url \@href}%
\providecommand \@href[1]{\@@startlink{#1}\@@href}%
\providecommand \@@href[1]{\endgroup#1\@@endlink}%
\providecommand \@sanitize@url [0]{\catcode `\\12\catcode `\$12\catcode
  `\&12\catcode `\#12\catcode `\^12\catcode `\_12\catcode `\%12\relax}%
\providecommand \@@startlink[1]{}%
\providecommand \@@endlink[0]{}%
\providecommand \url  [0]{\begingroup\@sanitize@url \@url }%
\providecommand \@url [1]{\endgroup\@href {#1}{\urlprefix }}%
\providecommand \urlprefix  [0]{URL }%
\providecommand \Eprint [0]{\href }%
\providecommand \doibase [0]{https://doi.org/}%
\providecommand \selectlanguage [0]{\@gobble}%
\providecommand \bibinfo  [0]{\@secondoftwo}%
\providecommand \bibfield  [0]{\@secondoftwo}%
\providecommand \translation [1]{[#1]}%
\providecommand \BibitemOpen [0]{}%
\providecommand \bibitemStop [0]{}%
\providecommand \bibitemNoStop [0]{.\EOS\space}%
\providecommand \EOS [0]{\spacefactor3000\relax}%
\providecommand \BibitemShut  [1]{\csname bibitem#1\endcsname}%
\let\auto@bib@innerbib\@empty
\bibitem [{Note1()}]{Note1}%
  \BibitemOpen
  \bibinfo {note} {Here we use ``chiral phonons'' to refer to circularly
  rotating phonon modes. Strictly speaking, circularly rotating phonon modes
  that are not propagating don’t have true chirality.}\BibitemShut {Stop}%
\bibitem [{\citenamefont {Grissonnanche}\ \emph {et~al.}(2019)\citenamefont
  {Grissonnanche}, \citenamefont {Legros}, \citenamefont {Badoux},
  \citenamefont {Lefran{\c{c}}ois}, \citenamefont {Zatko}, \citenamefont
  {Lizaire}, \citenamefont {Lalibert{\'e}}, \citenamefont {Gourgout},
  \citenamefont {Zhou}, \citenamefont {Pyon} \emph
  {et~al.}}]{grissonnanche2019giant}%
  \BibitemOpen
  \bibfield  {author} {\bibinfo {author} {\bibfnamefont {G.}~\bibnamefont
  {Grissonnanche}}, \bibinfo {author} {\bibfnamefont {A.}~\bibnamefont
  {Legros}}, \bibinfo {author} {\bibfnamefont {S.}~\bibnamefont {Badoux}},
  \bibinfo {author} {\bibfnamefont {E.}~\bibnamefont {Lefran{\c{c}}ois}},
  \bibinfo {author} {\bibfnamefont {V.}~\bibnamefont {Zatko}}, \bibinfo
  {author} {\bibfnamefont {M.}~\bibnamefont {Lizaire}}, \bibinfo {author}
  {\bibfnamefont {F.}~\bibnamefont {Lalibert{\'e}}}, \bibinfo {author}
  {\bibfnamefont {A.}~\bibnamefont {Gourgout}}, \bibinfo {author}
  {\bibfnamefont {J.-S.}\ \bibnamefont {Zhou}}, \bibinfo {author}
  {\bibfnamefont {S.}~\bibnamefont {Pyon}}, \emph {et~al.},\ }\bibfield
  {title} {\bibinfo {title} {Giant thermal hall conductivity in the pseudogap
  phase of cuprate superconductors},\ }\href@noop {} {\bibfield  {journal}
  {\bibinfo  {journal} {Nature}\ }\textbf {\bibinfo {volume} {571}},\ \bibinfo
  {pages} {376} (\bibinfo {year} {2019})}\BibitemShut {NoStop}%
\bibitem [{\citenamefont {Li}\ \emph {et~al.}(2020)\citenamefont {Li},
  \citenamefont {Fauqu{\'e}}, \citenamefont {Zhu},\ and\ \citenamefont
  {Behnia}}]{li2020phonon}%
  \BibitemOpen
  \bibfield  {author} {\bibinfo {author} {\bibfnamefont {X.}~\bibnamefont
  {Li}}, \bibinfo {author} {\bibfnamefont {B.}~\bibnamefont {Fauqu{\'e}}},
  \bibinfo {author} {\bibfnamefont {Z.}~\bibnamefont {Zhu}},\ and\ \bibinfo
  {author} {\bibfnamefont {K.}~\bibnamefont {Behnia}},\ }\bibfield  {title}
  {\bibinfo {title} {Phonon thermal hall effect in strontium titanate},\
  }\href@noop {} {\bibfield  {journal} {\bibinfo  {journal} {Phys. Rev. Lett.}\
  }\textbf {\bibinfo {volume} {124}},\ \bibinfo {pages} {105901} (\bibinfo
  {year} {2020})}\BibitemShut {NoStop}%
\bibitem [{\citenamefont {Zhang}\ and\ \citenamefont
  {Niu}(2014)}]{zhang2014angular}%
  \BibitemOpen
  \bibfield  {author} {\bibinfo {author} {\bibfnamefont {L.}~\bibnamefont
  {Zhang}}\ and\ \bibinfo {author} {\bibfnamefont {Q.}~\bibnamefont {Niu}},\
  }\bibfield  {title} {\bibinfo {title} {Angular momentum of phonons and the
  einstein--de haas effect},\ }\href@noop {} {\bibfield  {journal} {\bibinfo
  {journal} {Phys. Rev. Lett.}\ }\textbf {\bibinfo {volume} {112}},\ \bibinfo
  {pages} {085503} (\bibinfo {year} {2014})}\BibitemShut {NoStop}%
\bibitem [{\citenamefont {Dornes}\ \emph {et~al.}(2019)\citenamefont {Dornes},
  \citenamefont {Acremann}, \citenamefont {Savoini}, \citenamefont {Kubli},
  \citenamefont {Neugebauer}, \citenamefont {Abreu}, \citenamefont {Huber},
  \citenamefont {Lantz}, \citenamefont {Vaz}, \citenamefont {Lemke} \emph
  {et~al.}}]{dornes2019ultrafast}%
  \BibitemOpen
  \bibfield  {author} {\bibinfo {author} {\bibfnamefont {C.}~\bibnamefont
  {Dornes}}, \bibinfo {author} {\bibfnamefont {Y.}~\bibnamefont {Acremann}},
  \bibinfo {author} {\bibfnamefont {M.}~\bibnamefont {Savoini}}, \bibinfo
  {author} {\bibfnamefont {M.}~\bibnamefont {Kubli}}, \bibinfo {author}
  {\bibfnamefont {M.~J.}\ \bibnamefont {Neugebauer}}, \bibinfo {author}
  {\bibfnamefont {E.}~\bibnamefont {Abreu}}, \bibinfo {author} {\bibfnamefont
  {L.}~\bibnamefont {Huber}}, \bibinfo {author} {\bibfnamefont
  {G.}~\bibnamefont {Lantz}}, \bibinfo {author} {\bibfnamefont {C.~A.}\
  \bibnamefont {Vaz}}, \bibinfo {author} {\bibfnamefont {H.}~\bibnamefont
  {Lemke}}, \emph {et~al.},\ }\bibfield  {title} {\bibinfo {title} {The
  ultrafast einstein--de haas effect},\ }\href@noop {} {\bibfield  {journal}
  {\bibinfo  {journal} {Nature}\ }\textbf {\bibinfo {volume} {565}},\ \bibinfo
  {pages} {209} (\bibinfo {year} {2019})}\BibitemShut {NoStop}%
\bibitem [{\citenamefont {Zhang}\ and\ \citenamefont
  {Niu}(2015)}]{zhang2015chiral}%
  \BibitemOpen
  \bibfield  {author} {\bibinfo {author} {\bibfnamefont {L.}~\bibnamefont
  {Zhang}}\ and\ \bibinfo {author} {\bibfnamefont {Q.}~\bibnamefont {Niu}},\
  }\bibfield  {title} {\bibinfo {title} {Chiral phonons at high-symmetry points
  in monolayer hexagonal lattices},\ }\href@noop {} {\bibfield  {journal}
  {\bibinfo  {journal} {Phys. Rev. Lett.}\ }\textbf {\bibinfo {volume} {115}},\
  \bibinfo {pages} {115502} (\bibinfo {year} {2015})}\BibitemShut {NoStop}%
\bibitem [{\citenamefont {Zhu}\ \emph {et~al.}(2018)\citenamefont {Zhu},
  \citenamefont {Yi}, \citenamefont {Li}, \citenamefont {Xiao}, \citenamefont
  {Zhang}, \citenamefont {Yang}, \citenamefont {Kaindl}, \citenamefont {Li},
  \citenamefont {Wang},\ and\ \citenamefont {Zhang}}]{zhu2018observation}%
  \BibitemOpen
  \bibfield  {author} {\bibinfo {author} {\bibfnamefont {H.}~\bibnamefont
  {Zhu}}, \bibinfo {author} {\bibfnamefont {J.}~\bibnamefont {Yi}}, \bibinfo
  {author} {\bibfnamefont {M.-Y.}\ \bibnamefont {Li}}, \bibinfo {author}
  {\bibfnamefont {J.}~\bibnamefont {Xiao}}, \bibinfo {author} {\bibfnamefont
  {L.}~\bibnamefont {Zhang}}, \bibinfo {author} {\bibfnamefont {C.-W.}\
  \bibnamefont {Yang}}, \bibinfo {author} {\bibfnamefont {R.~A.}\ \bibnamefont
  {Kaindl}}, \bibinfo {author} {\bibfnamefont {L.-J.}\ \bibnamefont {Li}},
  \bibinfo {author} {\bibfnamefont {Y.}~\bibnamefont {Wang}},\ and\ \bibinfo
  {author} {\bibfnamefont {X.}~\bibnamefont {Zhang}},\ }\bibfield  {title}
  {\bibinfo {title} {Observation of chiral phonons},\ }\href@noop {} {\bibfield
   {journal} {\bibinfo  {journal} {Science}\ }\textbf {\bibinfo {volume}
  {359}},\ \bibinfo {pages} {579} (\bibinfo {year} {2018})}\BibitemShut
  {NoStop}%
\bibitem [{\citenamefont {Li}\ \emph {et~al.}(2019)\citenamefont {Li},
  \citenamefont {Wang}, \citenamefont {Jin}, \citenamefont {Lu}, \citenamefont
  {Lian}, \citenamefont {Meng}, \citenamefont {Blei}, \citenamefont {Gao},
  \citenamefont {Taniguchi}, \citenamefont {Watanabe} \emph
  {et~al.}}]{li2019emerging}%
  \BibitemOpen
  \bibfield  {author} {\bibinfo {author} {\bibfnamefont {Z.}~\bibnamefont
  {Li}}, \bibinfo {author} {\bibfnamefont {T.}~\bibnamefont {Wang}}, \bibinfo
  {author} {\bibfnamefont {C.}~\bibnamefont {Jin}}, \bibinfo {author}
  {\bibfnamefont {Z.}~\bibnamefont {Lu}}, \bibinfo {author} {\bibfnamefont
  {Z.}~\bibnamefont {Lian}}, \bibinfo {author} {\bibfnamefont {Y.}~\bibnamefont
  {Meng}}, \bibinfo {author} {\bibfnamefont {M.}~\bibnamefont {Blei}}, \bibinfo
  {author} {\bibfnamefont {S.}~\bibnamefont {Gao}}, \bibinfo {author}
  {\bibfnamefont {T.}~\bibnamefont {Taniguchi}}, \bibinfo {author}
  {\bibfnamefont {K.}~\bibnamefont {Watanabe}}, \emph {et~al.},\ }\bibfield
  {title} {\bibinfo {title} {Emerging photoluminescence from the dark-exciton
  phonon replica in monolayer {WSe}$_{2}$},\ }\href@noop {} {\bibfield
  {journal} {\bibinfo  {journal} {Nat. Commun.}\ }\textbf {\bibinfo {volume}
  {10}},\ \bibinfo {pages} {2469} (\bibinfo {year} {2019})}\BibitemShut
  {NoStop}%
\bibitem [{\citenamefont {He}\ \emph {et~al.}(2020)\citenamefont {He},
  \citenamefont {Rivera}, \citenamefont {Van~Tuan}, \citenamefont {Wilson},
  \citenamefont {Yang}, \citenamefont {Taniguchi}, \citenamefont {Watanabe},
  \citenamefont {Yan}, \citenamefont {Mandrus}, \citenamefont {Yu} \emph
  {et~al.}}]{he2020valley}%
  \BibitemOpen
  \bibfield  {author} {\bibinfo {author} {\bibfnamefont {M.}~\bibnamefont
  {He}}, \bibinfo {author} {\bibfnamefont {P.}~\bibnamefont {Rivera}}, \bibinfo
  {author} {\bibfnamefont {D.}~\bibnamefont {Van~Tuan}}, \bibinfo {author}
  {\bibfnamefont {N.~P.}\ \bibnamefont {Wilson}}, \bibinfo {author}
  {\bibfnamefont {M.}~\bibnamefont {Yang}}, \bibinfo {author} {\bibfnamefont
  {T.}~\bibnamefont {Taniguchi}}, \bibinfo {author} {\bibfnamefont
  {K.}~\bibnamefont {Watanabe}}, \bibinfo {author} {\bibfnamefont
  {J.}~\bibnamefont {Yan}}, \bibinfo {author} {\bibfnamefont {D.~G.}\
  \bibnamefont {Mandrus}}, \bibinfo {author} {\bibfnamefont {H.}~\bibnamefont
  {Yu}}, \emph {et~al.},\ }\bibfield  {title} {\bibinfo {title} {Valley phonons
  and exciton complexes in a monolayer semiconductor},\ }\href@noop {}
  {\bibfield  {journal} {\bibinfo  {journal} {Nat. Commun.}\ }\textbf {\bibinfo
  {volume} {11}},\ \bibinfo {pages} {618} (\bibinfo {year} {2020})}\BibitemShut
  {NoStop}%
\bibitem [{\citenamefont {Nova}\ \emph {et~al.}(2017)\citenamefont {Nova},
  \citenamefont {Cartella}, \citenamefont {Cantaluppi}, \citenamefont
  {F{\"o}rst}, \citenamefont {Bossini}, \citenamefont {Mikhaylovskiy},
  \citenamefont {Kimel}, \citenamefont {Merlin},\ and\ \citenamefont
  {Cavalleri}}]{nova2017effective}%
  \BibitemOpen
  \bibfield  {author} {\bibinfo {author} {\bibfnamefont {T.~F.}\ \bibnamefont
  {Nova}}, \bibinfo {author} {\bibfnamefont {A.}~\bibnamefont {Cartella}},
  \bibinfo {author} {\bibfnamefont {A.}~\bibnamefont {Cantaluppi}}, \bibinfo
  {author} {\bibfnamefont {M.}~\bibnamefont {F{\"o}rst}}, \bibinfo {author}
  {\bibfnamefont {D.}~\bibnamefont {Bossini}}, \bibinfo {author} {\bibfnamefont
  {R.~V.}\ \bibnamefont {Mikhaylovskiy}}, \bibinfo {author} {\bibfnamefont
  {A.~V.}\ \bibnamefont {Kimel}}, \bibinfo {author} {\bibfnamefont
  {R.}~\bibnamefont {Merlin}},\ and\ \bibinfo {author} {\bibfnamefont
  {A.}~\bibnamefont {Cavalleri}},\ }\bibfield  {title} {\bibinfo {title} {An
  effective magnetic field from optically driven phonons},\ }\href@noop {}
  {\bibfield  {journal} {\bibinfo  {journal} {Nat. Phys.}\ }\textbf {\bibinfo
  {volume} {13}},\ \bibinfo {pages} {132} (\bibinfo {year} {2017})}\BibitemShut
  {NoStop}%
\bibitem [{\citenamefont {Juraschek}\ \emph {et~al.}(2017)\citenamefont
  {Juraschek}, \citenamefont {Fechner}, \citenamefont {Balatsky},\ and\
  \citenamefont {Spaldin}}]{juraschek2017dynamical}%
  \BibitemOpen
  \bibfield  {author} {\bibinfo {author} {\bibfnamefont {D.~M.}\ \bibnamefont
  {Juraschek}}, \bibinfo {author} {\bibfnamefont {M.}~\bibnamefont {Fechner}},
  \bibinfo {author} {\bibfnamefont {A.~V.}\ \bibnamefont {Balatsky}},\ and\
  \bibinfo {author} {\bibfnamefont {N.~A.}\ \bibnamefont {Spaldin}},\
  }\bibfield  {title} {\bibinfo {title} {Dynamical multiferroicity},\
  }\href@noop {} {\bibfield  {journal} {\bibinfo  {journal} {Phys. Rev.
  Mater.}\ }\textbf {\bibinfo {volume} {1}},\ \bibinfo {pages} {014401}
  (\bibinfo {year} {2017})}\BibitemShut {NoStop}%
\bibitem [{\citenamefont {Cheng}\ \emph {et~al.}(2020)\citenamefont {Cheng},
  \citenamefont {Schumann}, \citenamefont {Wang}, \citenamefont {Zhang},
  \citenamefont {Barbalas}, \citenamefont {Stemmer},\ and\ \citenamefont
  {Armitage}}]{cheng2020large}%
  \BibitemOpen
  \bibfield  {author} {\bibinfo {author} {\bibfnamefont {B.}~\bibnamefont
  {Cheng}}, \bibinfo {author} {\bibfnamefont {T.}~\bibnamefont {Schumann}},
  \bibinfo {author} {\bibfnamefont {Y.}~\bibnamefont {Wang}}, \bibinfo {author}
  {\bibfnamefont {X.}~\bibnamefont {Zhang}}, \bibinfo {author} {\bibfnamefont
  {D.}~\bibnamefont {Barbalas}}, \bibinfo {author} {\bibfnamefont
  {S.}~\bibnamefont {Stemmer}},\ and\ \bibinfo {author} {\bibfnamefont
  {N.}~\bibnamefont {Armitage}},\ }\bibfield  {title} {\bibinfo {title} {A
  large effective phonon magnetic moment in a dirac semimetal},\ }\href@noop {}
  {\bibfield  {journal} {\bibinfo  {journal} {Nano Lett.}\ }\textbf {\bibinfo
  {volume} {20}},\ \bibinfo {pages} {5991} (\bibinfo {year}
  {2020})}\BibitemShut {NoStop}%
\bibitem [{\citenamefont {Baydin}\ \emph {et~al.}(2022)\citenamefont {Baydin},
  \citenamefont {Hernandez}, \citenamefont {Rodriguez-Vega}, \citenamefont
  {Okazaki}, \citenamefont {Tay}, \citenamefont {G~Timothy~Noe}, \citenamefont
  {Katayama}, \citenamefont {Takeda}, \citenamefont {Nojiri}, \citenamefont
  {Rappl} \emph {et~al.}}]{baydin2022magnetic}%
  \BibitemOpen
  \bibfield  {author} {\bibinfo {author} {\bibfnamefont {A.}~\bibnamefont
  {Baydin}}, \bibinfo {author} {\bibfnamefont {F.~G.}\ \bibnamefont
  {Hernandez}}, \bibinfo {author} {\bibfnamefont {M.}~\bibnamefont
  {Rodriguez-Vega}}, \bibinfo {author} {\bibfnamefont {A.~K.}\ \bibnamefont
  {Okazaki}}, \bibinfo {author} {\bibfnamefont {F.}~\bibnamefont {Tay}},
  \bibinfo {author} {\bibfnamefont {I.}~\bibnamefont {G~Timothy~Noe}}, \bibinfo
  {author} {\bibfnamefont {I.}~\bibnamefont {Katayama}}, \bibinfo {author}
  {\bibfnamefont {J.}~\bibnamefont {Takeda}}, \bibinfo {author} {\bibfnamefont
  {H.}~\bibnamefont {Nojiri}}, \bibinfo {author} {\bibfnamefont {P.~H.}\
  \bibnamefont {Rappl}}, \emph {et~al.},\ }\bibfield  {title} {\bibinfo {title}
  {Magnetic control of soft chiral phonons in {PbTe}},\ }\href@noop {}
  {\bibfield  {journal} {\bibinfo  {journal} {Phys. Rev. Lett.}\ }\textbf
  {\bibinfo {volume} {128}},\ \bibinfo {pages} {075901} (\bibinfo {year}
  {2022})}\BibitemShut {NoStop}%
\bibitem [{\citenamefont {Hernandez}\ \emph {et~al.}(2022)\citenamefont
  {Hernandez}, \citenamefont {Baydin}, \citenamefont {Chaudhary}, \citenamefont
  {Tay}, \citenamefont {Katayama}, \citenamefont {Takeda}, \citenamefont
  {Nojiri}, \citenamefont {Okazaki}, \citenamefont {Rappl}, \citenamefont
  {Abramof} \emph {et~al.}}]{hernandez2022chiral}%
  \BibitemOpen
  \bibfield  {author} {\bibinfo {author} {\bibfnamefont {F.~G.}\ \bibnamefont
  {Hernandez}}, \bibinfo {author} {\bibfnamefont {A.}~\bibnamefont {Baydin}},
  \bibinfo {author} {\bibfnamefont {S.}~\bibnamefont {Chaudhary}}, \bibinfo
  {author} {\bibfnamefont {F.}~\bibnamefont {Tay}}, \bibinfo {author}
  {\bibfnamefont {I.}~\bibnamefont {Katayama}}, \bibinfo {author}
  {\bibfnamefont {J.}~\bibnamefont {Takeda}}, \bibinfo {author} {\bibfnamefont
  {H.}~\bibnamefont {Nojiri}}, \bibinfo {author} {\bibfnamefont {A.~K.}\
  \bibnamefont {Okazaki}}, \bibinfo {author} {\bibfnamefont {P.~H.}\
  \bibnamefont {Rappl}}, \bibinfo {author} {\bibfnamefont {E.}~\bibnamefont
  {Abramof}}, \emph {et~al.},\ }\bibfield  {title} {\bibinfo {title} {Chiral
  phonons with giant magnetic moments in a topological crystalline insulator},\
  }\href@noop {} {\bibfield  {journal} {\bibinfo  {journal} {arXiv preprint
  arXiv:2208.12235}\ } (\bibinfo {year} {2022})}\BibitemShut {NoStop}%
\bibitem [{\citenamefont {Basini}\ \emph {et~al.}(2022)\citenamefont {Basini},
  \citenamefont {Pancaldi}, \citenamefont {Wehinger}, \citenamefont {Udina},
  \citenamefont {Tadano}, \citenamefont {Hoffmann}, \citenamefont {Balatsky},\
  and\ \citenamefont {Bonetti}}]{basini2022terahertz}%
  \BibitemOpen
  \bibfield  {author} {\bibinfo {author} {\bibfnamefont {M.}~\bibnamefont
  {Basini}}, \bibinfo {author} {\bibfnamefont {M.}~\bibnamefont {Pancaldi}},
  \bibinfo {author} {\bibfnamefont {B.}~\bibnamefont {Wehinger}}, \bibinfo
  {author} {\bibfnamefont {M.}~\bibnamefont {Udina}}, \bibinfo {author}
  {\bibfnamefont {T.}~\bibnamefont {Tadano}}, \bibinfo {author} {\bibfnamefont
  {M.}~\bibnamefont {Hoffmann}}, \bibinfo {author} {\bibfnamefont
  {A.}~\bibnamefont {Balatsky}},\ and\ \bibinfo {author} {\bibfnamefont
  {S.}~\bibnamefont {Bonetti}},\ }\bibfield  {title} {\bibinfo {title}
  {Terahertz electric-field driven dynamical multiferroicity in {SrTiO}$_3$},\
  }\href@noop {} {\bibfield  {journal} {\bibinfo  {journal} {arXiv preprint
  arXiv:2210.01690}\ } (\bibinfo {year} {2022})}\BibitemShut {NoStop}%
\bibitem [{\citenamefont {Juraschek}\ and\ \citenamefont
  {Spaldin}(2019)}]{juraschek2019orbital}%
  \BibitemOpen
  \bibfield  {author} {\bibinfo {author} {\bibfnamefont {D.~M.}\ \bibnamefont
  {Juraschek}}\ and\ \bibinfo {author} {\bibfnamefont {N.~A.}\ \bibnamefont
  {Spaldin}},\ }\bibfield  {title} {\bibinfo {title} {Orbital magnetic moments
  of phonons},\ }\href@noop {} {\bibfield  {journal} {\bibinfo  {journal}
  {Phys. Rev. Mater.}\ }\textbf {\bibinfo {volume} {3}},\ \bibinfo {pages}
  {064405} (\bibinfo {year} {2019})}\BibitemShut {NoStop}%
\bibitem [{\citenamefont {Ceresoli}()}]{ceresoli2002berry}%
  \BibitemOpen
  \bibfield  {author} {\bibinfo {author} {\bibfnamefont {D.}~\bibnamefont
  {Ceresoli}},\ }\bibfield  {title} {\bibinfo {title} {Berry phase calculations
  of the rotational and pseudorotational g-factor in molecules and solids},\
  }\href@noop {} {\bibinfo  {journal} {Ph.D thesis, 2002}\ }\BibitemShut
  {NoStop}%
\bibitem [{\citenamefont {Ceresoli}\ and\ \citenamefont
  {Tosatti}(2002)}]{ceresoli2002PRL}%
  \BibitemOpen
\bibfield  {journal} {  }\bibfield  {author} {\bibinfo {author} {\bibfnamefont
  {D.}~\bibnamefont {Ceresoli}}\ and\ \bibinfo {author} {\bibfnamefont
  {E.}~\bibnamefont {Tosatti}},\ }\bibfield  {title} {\bibinfo {title}
  {Berry-phase calculation of magnetic screening and rotational g factor in
  molecules and solids},\ }\href@noop {} {\bibfield  {journal} {\bibinfo
  {journal} {Phys. Rev. Lett.}\ }\textbf {\bibinfo {volume} {89}},\ \bibinfo
  {pages} {116402} (\bibinfo {year} {2002})}\BibitemShut {NoStop}%
\bibitem [{\citenamefont {Zabalo}\ \emph {et~al.}(2022)\citenamefont {Zabalo},
  \citenamefont {Dreyer},\ and\ \citenamefont
  {Stengel}}]{zabalo2022rotational}%
  \BibitemOpen
  \bibfield  {author} {\bibinfo {author} {\bibfnamefont {A.}~\bibnamefont
  {Zabalo}}, \bibinfo {author} {\bibfnamefont {C.~E.}\ \bibnamefont {Dreyer}},\
  and\ \bibinfo {author} {\bibfnamefont {M.}~\bibnamefont {Stengel}},\
  }\bibfield  {title} {\bibinfo {title} {Rotational g factors and lorentz
  forces of molecules and solids from density functional perturbation theory},\
  }\href@noop {} {\bibfield  {journal} {\bibinfo  {journal} {Phys. Rev. B}\
  }\textbf {\bibinfo {volume} {105}},\ \bibinfo {pages} {094305} (\bibinfo
  {year} {2022})}\BibitemShut {NoStop}%
\bibitem [{\citenamefont {Dong}\ and\ \citenamefont
  {Niu}(2018)}]{dong2018geometrodynamics}%
  \BibitemOpen
  \bibfield  {author} {\bibinfo {author} {\bibfnamefont {L.}~\bibnamefont
  {Dong}}\ and\ \bibinfo {author} {\bibfnamefont {Q.}~\bibnamefont {Niu}},\
  }\bibfield  {title} {\bibinfo {title} {Geometrodynamics of electrons in a
  crystal under position and time-dependent deformation},\ }\href@noop {}
  {\bibfield  {journal} {\bibinfo  {journal} {Phys. Rev. B}\ }\textbf {\bibinfo
  {volume} {98}},\ \bibinfo {pages} {115162} (\bibinfo {year}
  {2018})}\BibitemShut {NoStop}%
\bibitem [{\citenamefont {Trifunovic}\ \emph {et~al.}(2019)\citenamefont
  {Trifunovic}, \citenamefont {Ono},\ and\ \citenamefont
  {Watanabe}}]{trifunovic2019geometric}%
  \BibitemOpen
  \bibfield  {author} {\bibinfo {author} {\bibfnamefont {L.}~\bibnamefont
  {Trifunovic}}, \bibinfo {author} {\bibfnamefont {S.}~\bibnamefont {Ono}},\
  and\ \bibinfo {author} {\bibfnamefont {H.}~\bibnamefont {Watanabe}},\
  }\bibfield  {title} {\bibinfo {title} {Geometric orbital magnetization in
  adiabatic processes},\ }\href@noop {} {\bibfield  {journal} {\bibinfo
  {journal} {Phys. Rev. B}\ }\textbf {\bibinfo {volume} {100}},\ \bibinfo
  {pages} {054408} (\bibinfo {year} {2019})}\BibitemShut {NoStop}%
\bibitem [{\citenamefont {Xiao}\ \emph {et~al.}(2021)\citenamefont {Xiao},
  \citenamefont {Ren},\ and\ \citenamefont {Xiong}}]{xiao2021adiabatically}%
  \BibitemOpen
  \bibfield  {author} {\bibinfo {author} {\bibfnamefont {C.}~\bibnamefont
  {Xiao}}, \bibinfo {author} {\bibfnamefont {Y.}~\bibnamefont {Ren}},\ and\
  \bibinfo {author} {\bibfnamefont {B.}~\bibnamefont {Xiong}},\ }\bibfield
  {title} {\bibinfo {title} {Adiabatically induced orbital magnetization},\
  }\href@noop {} {\bibfield  {journal} {\bibinfo  {journal} {Phys. Rev. B}\
  }\textbf {\bibinfo {volume} {103}},\ \bibinfo {pages} {115432} (\bibinfo
  {year} {2021})}\BibitemShut {NoStop}%
\bibitem [{\citenamefont {Ren}\ \emph {et~al.}(2021)\citenamefont {Ren},
  \citenamefont {Xiao}, \citenamefont {Saparov},\ and\ \citenamefont
  {Niu}}]{ren2021phonon}%
  \BibitemOpen
  \bibfield  {author} {\bibinfo {author} {\bibfnamefont {Y.}~\bibnamefont
  {Ren}}, \bibinfo {author} {\bibfnamefont {C.}~\bibnamefont {Xiao}}, \bibinfo
  {author} {\bibfnamefont {D.}~\bibnamefont {Saparov}},\ and\ \bibinfo {author}
  {\bibfnamefont {Q.}~\bibnamefont {Niu}},\ }\bibfield  {title} {\bibinfo
  {title} {Phonon magnetic moment from electronic topological magnetization},\
  }\href@noop {} {\bibfield  {journal} {\bibinfo  {journal} {Phys. Rev. Lett.}\
  }\textbf {\bibinfo {volume} {127}},\ \bibinfo {pages} {186403} (\bibinfo
  {year} {2021})}\BibitemShut {NoStop}%
\bibitem [{sup()}]{supp}%
  \BibitemOpen
  \href@noop {} {\bibinfo  {journal} {See Supplemental Materials for the
  calculation method and numerical details of phonon magnetic moments}\
  }\BibitemShut {NoStop}%
\bibitem [{\citenamefont {Sai}\ \emph {et~al.}(2002)\citenamefont {Sai},
  \citenamefont {Rabe},\ and\ \citenamefont {Vanderbilt}}]{sai2002theory}%
  \BibitemOpen
\bibfield  {journal} {  }\bibfield  {author} {\bibinfo {author} {\bibfnamefont
  {N.}~\bibnamefont {Sai}}, \bibinfo {author} {\bibfnamefont {K.~M.}\
  \bibnamefont {Rabe}},\ and\ \bibinfo {author} {\bibfnamefont
  {D.}~\bibnamefont {Vanderbilt}},\ }\bibfield  {title} {\bibinfo {title}
  {Theory of structural response to macroscopic electric fields in
  ferroelectric systems},\ }\href@noop {} {\bibfield  {journal} {\bibinfo
  {journal} {Phys. Rev. B}\ }\textbf {\bibinfo {volume} {66}},\ \bibinfo
  {pages} {104108} (\bibinfo {year} {2002})}\BibitemShut {NoStop}%
\bibitem [{\citenamefont {Souza}\ \emph {et~al.}(2004)\citenamefont {Souza},
  \citenamefont {{\'I}niguez},\ and\ \citenamefont
  {Vanderbilt}}]{souza2004dynamics}%
  \BibitemOpen
  \bibfield  {author} {\bibinfo {author} {\bibfnamefont {I.}~\bibnamefont
  {Souza}}, \bibinfo {author} {\bibfnamefont {J.}~\bibnamefont {{\'I}niguez}},\
  and\ \bibinfo {author} {\bibfnamefont {D.}~\bibnamefont {Vanderbilt}},\
  }\bibfield  {title} {\bibinfo {title} {Dynamics of berry-phase polarization
  in time-dependent electric fields},\ }\href@noop {} {\bibfield  {journal}
  {\bibinfo  {journal} {Phys. Rev. B}\ }\textbf {\bibinfo {volume} {69}},\
  \bibinfo {pages} {085106} (\bibinfo {year} {2004})}\BibitemShut {NoStop}%
\bibitem [{\citenamefont {Wang}\ \emph {et~al.}(2006)\citenamefont {Wang},
  \citenamefont {Yates}, \citenamefont {Souza},\ and\ \citenamefont
  {Vanderbilt}}]{wang2006ab}%
  \BibitemOpen
  \bibfield  {author} {\bibinfo {author} {\bibfnamefont {X.}~\bibnamefont
  {Wang}}, \bibinfo {author} {\bibfnamefont {J.~R.}\ \bibnamefont {Yates}},
  \bibinfo {author} {\bibfnamefont {I.}~\bibnamefont {Souza}},\ and\ \bibinfo
  {author} {\bibfnamefont {D.}~\bibnamefont {Vanderbilt}},\ }\bibfield  {title}
  {\bibinfo {title} {Ab initio calculation of the anomalous hall conductivity
  by wannier interpolation},\ }\href@noop {} {\bibfield  {journal} {\bibinfo
  {journal} {Phys. Rev. B}\ }\textbf {\bibinfo {volume} {74}},\ \bibinfo
  {pages} {195118} (\bibinfo {year} {2006})}\BibitemShut {NoStop}%
\bibitem [{\citenamefont {Giannozzi}\ \emph {et~al.}(2009)\citenamefont
  {Giannozzi}, \citenamefont {Baroni}, \citenamefont {Bonini}, \citenamefont
  {Calandra}, \citenamefont {Car}, \citenamefont {Cavazzoni}, \citenamefont
  {Ceresoli}, \citenamefont {Chiarotti}, \citenamefont {Cococcioni},
  \citenamefont {Dabo} \emph {et~al.}}]{giannozzi2009quantum}%
  \BibitemOpen
  \bibfield  {author} {\bibinfo {author} {\bibfnamefont {P.}~\bibnamefont
  {Giannozzi}}, \bibinfo {author} {\bibfnamefont {S.}~\bibnamefont {Baroni}},
  \bibinfo {author} {\bibfnamefont {N.}~\bibnamefont {Bonini}}, \bibinfo
  {author} {\bibfnamefont {M.}~\bibnamefont {Calandra}}, \bibinfo {author}
  {\bibfnamefont {R.}~\bibnamefont {Car}}, \bibinfo {author} {\bibfnamefont
  {C.}~\bibnamefont {Cavazzoni}}, \bibinfo {author} {\bibfnamefont
  {D.}~\bibnamefont {Ceresoli}}, \bibinfo {author} {\bibfnamefont {G.~L.}\
  \bibnamefont {Chiarotti}}, \bibinfo {author} {\bibfnamefont {M.}~\bibnamefont
  {Cococcioni}}, \bibinfo {author} {\bibfnamefont {I.}~\bibnamefont {Dabo}},
  \emph {et~al.},\ }\bibfield  {title} {\bibinfo {title} {Quantum espresso: a
  modular and open-source software project for quantum simulations of
  materials},\ }\href@noop {} {\bibfield  {journal} {\bibinfo  {journal} {J.
  Phys. Condens. Matter.}\ }\textbf {\bibinfo {volume} {21}},\ \bibinfo {pages}
  {395502} (\bibinfo {year} {2009})}\BibitemShut {NoStop}%
\bibitem [{\citenamefont {Hamann}(2013)}]{hamann2013optimized}%
  \BibitemOpen
  \bibfield  {author} {\bibinfo {author} {\bibfnamefont {D.}~\bibnamefont
  {Hamann}},\ }\bibfield  {title} {\bibinfo {title} {Optimized norm-conserving
  vanderbilt pseudopotentials},\ }\href@noop {} {\bibfield  {journal} {\bibinfo
   {journal} {Phys. Rev. B}\ }\textbf {\bibinfo {volume} {88}},\ \bibinfo
  {pages} {085117} (\bibinfo {year} {2013})}\BibitemShut {NoStop}%
\bibitem [{\citenamefont {Ramsey~Jr}(1940)}]{ramsey1940rotational}%
  \BibitemOpen
  \bibfield  {author} {\bibinfo {author} {\bibfnamefont {N.}~\bibnamefont
  {Ramsey~Jr}},\ }\bibfield  {title} {\bibinfo {title} {The rotational magnetic
  moments of {H}$_2$, {D}$_2$, and {HD} molecules. the rotational
  radiofrequency spectra of {H}$_2$, {D}$_2$, and {D} in magnetic fields},\
  }\href@noop {} {\bibfield  {journal} {\bibinfo  {journal} {Phys. Rev.}\
  }\textbf {\bibinfo {volume} {58}},\ \bibinfo {pages} {226} (\bibinfo {year}
  {1940})}\BibitemShut {NoStop}%
\bibitem [{\citenamefont {Hamada}\ and\ \citenamefont
  {Murakami}(2020)}]{hamada2020conversion}%
  \BibitemOpen
  \bibfield  {author} {\bibinfo {author} {\bibfnamefont {M.}~\bibnamefont
  {Hamada}}\ and\ \bibinfo {author} {\bibfnamefont {S.}~\bibnamefont
  {Murakami}},\ }\bibfield  {title} {\bibinfo {title} {Conversion between
  electron spin and microscopic atomic rotation},\ }\href@noop {} {\bibfield
  {journal} {\bibinfo  {journal} {Phys. Rev. Res.}\ }\textbf {\bibinfo {volume}
  {2}},\ \bibinfo {pages} {023275} (\bibinfo {year} {2020})}\BibitemShut
  {NoStop}%
\bibitem [{Note2()}]{Note2}%
  \BibitemOpen
  \bibinfo {note} {Ref.~\cite {ren2021phonon} considered the magnetic moment of
  the $K$-point phonons in gapped monolayer graphene. Here we consider the
  $\Gamma $-point phonons in a gated bilayer graphene}\BibitemShut {NoStop}%
\bibitem [{\citenamefont {Zhang}\ \emph {et~al.}(2009)\citenamefont {Zhang},
  \citenamefont {Tang}, \citenamefont {Girit}, \citenamefont {Hao},
  \citenamefont {Martin}, \citenamefont {Zettl}, \citenamefont {Crommie},
  \citenamefont {Shen},\ and\ \citenamefont {Wang}}]{zhang2009direct}%
  \BibitemOpen
  \bibfield  {author} {\bibinfo {author} {\bibfnamefont {Y.}~\bibnamefont
  {Zhang}}, \bibinfo {author} {\bibfnamefont {T.-T.}\ \bibnamefont {Tang}},
  \bibinfo {author} {\bibfnamefont {C.}~\bibnamefont {Girit}}, \bibinfo
  {author} {\bibfnamefont {Z.}~\bibnamefont {Hao}}, \bibinfo {author}
  {\bibfnamefont {M.~C.}\ \bibnamefont {Martin}}, \bibinfo {author}
  {\bibfnamefont {A.}~\bibnamefont {Zettl}}, \bibinfo {author} {\bibfnamefont
  {M.~F.}\ \bibnamefont {Crommie}}, \bibinfo {author} {\bibfnamefont {Y.~R.}\
  \bibnamefont {Shen}},\ and\ \bibinfo {author} {\bibfnamefont
  {F.}~\bibnamefont {Wang}},\ }\bibfield  {title} {\bibinfo {title} {Direct
  observation of a widely tunable bandgap in bilayer graphene},\ }\href@noop {}
  {\bibfield  {journal} {\bibinfo  {journal} {Nature}\ }\textbf {\bibinfo
  {volume} {459}},\ \bibinfo {pages} {820} (\bibinfo {year}
  {2009})}\BibitemShut {NoStop}%
\bibitem [{Note3()}]{Note3}%
  \BibitemOpen
  \bibinfo {note} {We note that the magnetic moment is invariant under spatial
  inversion, therefore the phonon magnetic moment can be nonzero even in
  covalent materials with inversion symmetry such as silicon and blue
  phosphorus listed in Table~\ref {tab}}\BibitemShut {NoStop}%
\bibitem [{\citenamefont {Schaack}(1976)}]{schaack1976observation}%
  \BibitemOpen
  \bibfield  {author} {\bibinfo {author} {\bibfnamefont {G.}~\bibnamefont
  {Schaack}},\ }\bibfield  {title} {\bibinfo {title} {Observation of circularly
  polarized phonon states in an external magnetic field},\ }\href@noop {}
  {\bibfield  {journal} {\bibinfo  {journal} {J. Phys. C: Solid State Phys.}\
  }\textbf {\bibinfo {volume} {9}},\ \bibinfo {pages} {L297} (\bibinfo {year}
  {1976})}\BibitemShut {NoStop}%
\bibitem [{\citenamefont {Schaack}(1977)}]{schaack1977magnetic}%
  \BibitemOpen
  \bibfield  {author} {\bibinfo {author} {\bibfnamefont {G.}~\bibnamefont
  {Schaack}},\ }\bibfield  {title} {\bibinfo {title} {Magnetic field dependent
  splitting of doubly degenerate phonon states in anhydrous
  cerium-trichloride},\ }\href@noop {} {\bibfield  {journal} {\bibinfo
  {journal} {Z. Phys. B}\ }\textbf {\bibinfo {volume} {26}},\ \bibinfo {pages}
  {49} (\bibinfo {year} {1977})}\BibitemShut {NoStop}%
\bibitem [{\citenamefont {Shin}\ \emph {et~al.}(2018)\citenamefont {Shin},
  \citenamefont {H{\"u}bener}, \citenamefont {De~Giovannini}, \citenamefont
  {Jin}, \citenamefont {Rubio},\ and\ \citenamefont {Park}}]{shin2018phonon}%
  \BibitemOpen
  \bibfield  {author} {\bibinfo {author} {\bibfnamefont {D.}~\bibnamefont
  {Shin}}, \bibinfo {author} {\bibfnamefont {H.}~\bibnamefont {H{\"u}bener}},
  \bibinfo {author} {\bibfnamefont {U.}~\bibnamefont {De~Giovannini}}, \bibinfo
  {author} {\bibfnamefont {H.}~\bibnamefont {Jin}}, \bibinfo {author}
  {\bibfnamefont {A.}~\bibnamefont {Rubio}},\ and\ \bibinfo {author}
  {\bibfnamefont {N.}~\bibnamefont {Park}},\ }\bibfield  {title} {\bibinfo
  {title} {Phonon-driven spin-floquet magneto-valleytronics in {MoS}$_2$},\
  }\href@noop {} {\bibfield  {journal} {\bibinfo  {journal} {Nat. Commun.}\
  }\textbf {\bibinfo {volume} {9}},\ \bibinfo {pages} {1} (\bibinfo {year}
  {2018})}\BibitemShut {NoStop}%
\bibitem [{\citenamefont {Schirhagl}\ \emph {et~al.}(2014)\citenamefont
  {Schirhagl}, \citenamefont {Chang}, \citenamefont {Loretz},\ and\
  \citenamefont {Degen}}]{schirhagl2014nitrogen}%
  \BibitemOpen
  \bibfield  {author} {\bibinfo {author} {\bibfnamefont {R.}~\bibnamefont
  {Schirhagl}}, \bibinfo {author} {\bibfnamefont {K.}~\bibnamefont {Chang}},
  \bibinfo {author} {\bibfnamefont {M.}~\bibnamefont {Loretz}},\ and\ \bibinfo
  {author} {\bibfnamefont {C.~L.}\ \bibnamefont {Degen}},\ }\bibfield  {title}
  {\bibinfo {title} {Nitrogen-vacancy centers in diamond: nanoscale sensors for
  physics and biology},\ }\href@noop {} {\bibfield  {journal} {\bibinfo
  {journal} {Annu. Rev. Phys. Chem}\ }\textbf {\bibinfo {volume} {65}},\
  \bibinfo {pages} {83} (\bibinfo {year} {2014})}\BibitemShut {NoStop}%
\end{thebibliography}
\end{document}